\documentclass[prl,showpacs,aps,twocolumn]{revtex4}%
\usepackage{amsfonts}%
\usepackage{amsmath}%
\usepackage{amssymb}%
\usepackage{graphicx}

\begin{document}
\preprint{ }

\bigskip
\title{Structure of many-body continuum states in the proximity of  exceptional points}

\author
{J. Oko{\l}owicz$^{1}$, and M. P{\l}oszajczak$^{2}$}

\affiliation{
$^{1}$ Institute of Nuclear Physics, Radzikowskiego 152, PL-31342 Krak\'ow, Poland\\
$^{2}$ Grand Acc\'{e}l\'{e}rateur National d'Ions Lourds (GANIL), CEA/DSM -- CNRS/IN2P3, BP 5027, F-14076 Caen Cedex 05, France \\
}

\date{\today}

\begin{abstract}
We demonstrate existence of exceptional points in many-body scattering continuum of atomic nucleus and  discuss their salient effects on the example of one-nucleon spectroscopic factors. 
\end{abstract}


\bigskip

\maketitle

\section{Introduction}
Degeneracies of eigenvalues of matrices dependent on parameters is a problem of great importance in Physics. The interest in this problem goes back to Hamilton \cite{Ham33} who pointed out 
a conical refraction due to coincident eigenvalues \cite{Ber99}. Later, the significance of degeneracies of matrix spectra has been revealed in quantum physics \cite{VNe29,Ber85}, crystal optics \cite{Her37} and chemistry \cite{Tel37}. Eigenvalue degeneracies are also essential for an understanding of the spectral fluctuations \cite{Ber81} and the onset of quantum chaos \cite{Ber85}. Among those degeneracies, one finds the exceptional points (EPs) \cite{Kat95,Zir83,Hei91} where two  Riemann sheets of the eigenvalues are entangled by the square-root type of singularity. The EPs are generic singularities and, hence, can be found in numerous situations, e.g. the resonance problems \cite{Mon96,Her06} or the quantum phase transitions \cite{Hei06,Cej07,Dor01}.  

Much effort has been devoted to studies of degeneracies associated with avoided crossings in quantal spectra, focusing mainly on the topological structure of  Hilbert space and the geometric phase 
\cite{Ber84,Lau94,Mon96}  which a quantum system acquires when transported adiabatically around the singularity in parameter space \cite{Ber84,Rev88}. These novel effects have attracted a considerable experimental interest as well \cite{Dem01}.

In many-body systems, EPs have been studied in schematic models, such as the Lipkin 
model \cite{Hei91a}, the interacting boson model \cite{Cej07} and, recently, the  3-level pairing model \cite{Duk08,Duk08a} which belongs to a general class of  the Richardson-Gaudin models 
\cite{Duk04,Ort05}. In the latter case, it has been found that eigenvalue crossings in the complex-extended parameter space of the Hamiltonian and the reduction in the number of EPs uniquely define the quantum integrable system \cite{Duk08}. If true, this conjecture opens a possibility for studies of fingerprints of the chaotic dynamics in the quantum regime for small dimensional Hilbert spaces.

In the case of closed quantum system (CQS) described by a hermitian Hamiltonian, mixing of two eigenvalues leads generically to the level repulsion and the  avoided level crossing. In the open quantum system (OQS), eigenvalues corresponding to unbound levels may exhibit not only crossing of either energies or widths but also true degeneracies due to a joint crossing of both energies and widths. The latter case corresponds to a degenerate double pole of the scattering matrix 
(S-matrix) \cite{Las66,Kyl98,Rot02,Zno06}. Hence, OQS is a natural system to look for singularities 
of the EP type. 

First attempt to search for EPs in the realistic many-body model of atomic nucleus has been made recently using the  real-energy continuum shell model \cite{Bar77}, the so-called Shell Model Embedded in the Continuum (SMEC) \cite{SMEC,Oko03,SMEC_2p}, which is defined in projected Hilbert spaces with at most two particles in the scattering continuum. The total function space of the $A$-particle system in SMEC consists of two sets: the set of square-integrable functions $\{\Phi_i^{A}\}$, used in the standard nuclear Shell Model (SM), and the set of continuum states (the environment of the CQS) 
$\{\zeta_E^{c(+)}\}$ embedding the SM. In one-particle continuum problems, channels $c$ are determined by the motion of an unbound particle in a state $l_j$ relative to the  $A-1$ nucleus in a certain SM state $\Phi_j^{A-1}$. An index '+' denotes an outgoing asymptotics of a continuum state 
$\zeta_E^{c(+)}$. Eigenstates of the OQS are found by solving the eigenvalue problem for an effective Hamiltonian:
\begin{eqnarray}
\label{heff}
H_{QQ}^{\mbox{\small eff}}(E) = H_{QQ} + H_{QP}G_{P}^{(+)}(E)H_{PQ} 
\end{eqnarray}
in the function space $\{\Phi_i^{A}\}$ of discrete states. Indices $Q$ and $P$ in (\ref{heff}) denote the projected Hilbert subspaces with 0 and 1 nucleon in the scattering continuum, respectively. 
$G_{P}^{(+)}(E)$~ is the Green function of a single nucleon in the $P$ subspace. Whereas the SM Hamiltonian ($H_{QQ}$)  is hermitian, $H_{QQ}^{\mbox{\small eff}}$ is an energy dependent, non-hermitian operator above the particle emission threshold and a hermitian operator below the emission threshold.  For each total angular momentum $J$ and parity $\pi$ separately, the $Q-P$ coupling term in (\ref{heff}) gives rise to an external mixing of different SM states $\{\Phi_k^{A}(J^{\pi})\}$ via the coupling to common decay channels, both closed and opened. The amount of this external mixing depends on the position of considered SM states with respect to the decay channel thresholds \cite{Cha06}.
More details about the SMEC approach and its recent applications can be found in Refs. 
\cite{Oko03,Oko09}.

\begin{figure}[hbt]
\begin{center}
\includegraphics[width=5.5cm,angle=-90]{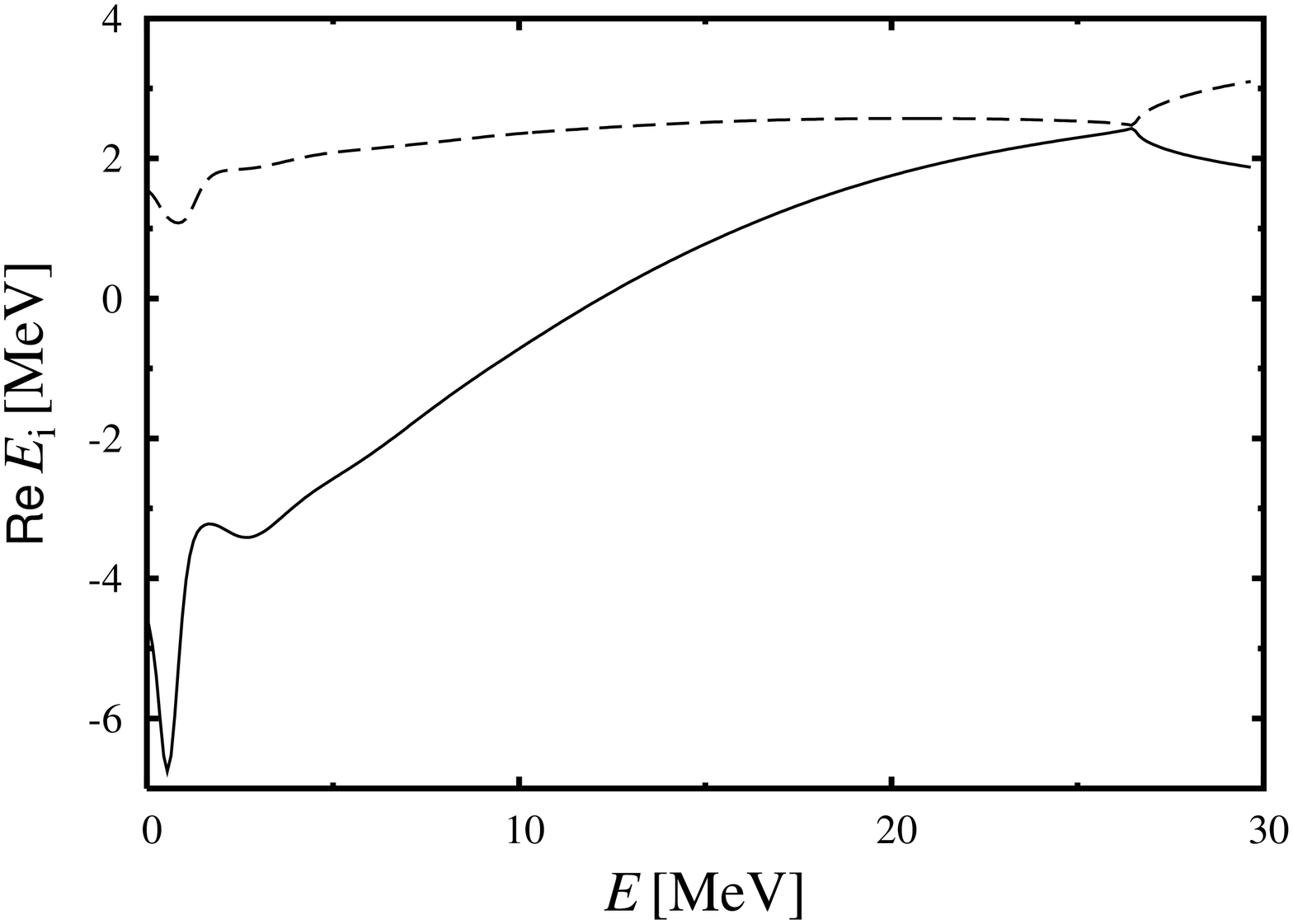}
\includegraphics[width=5.5cm,angle=-90]{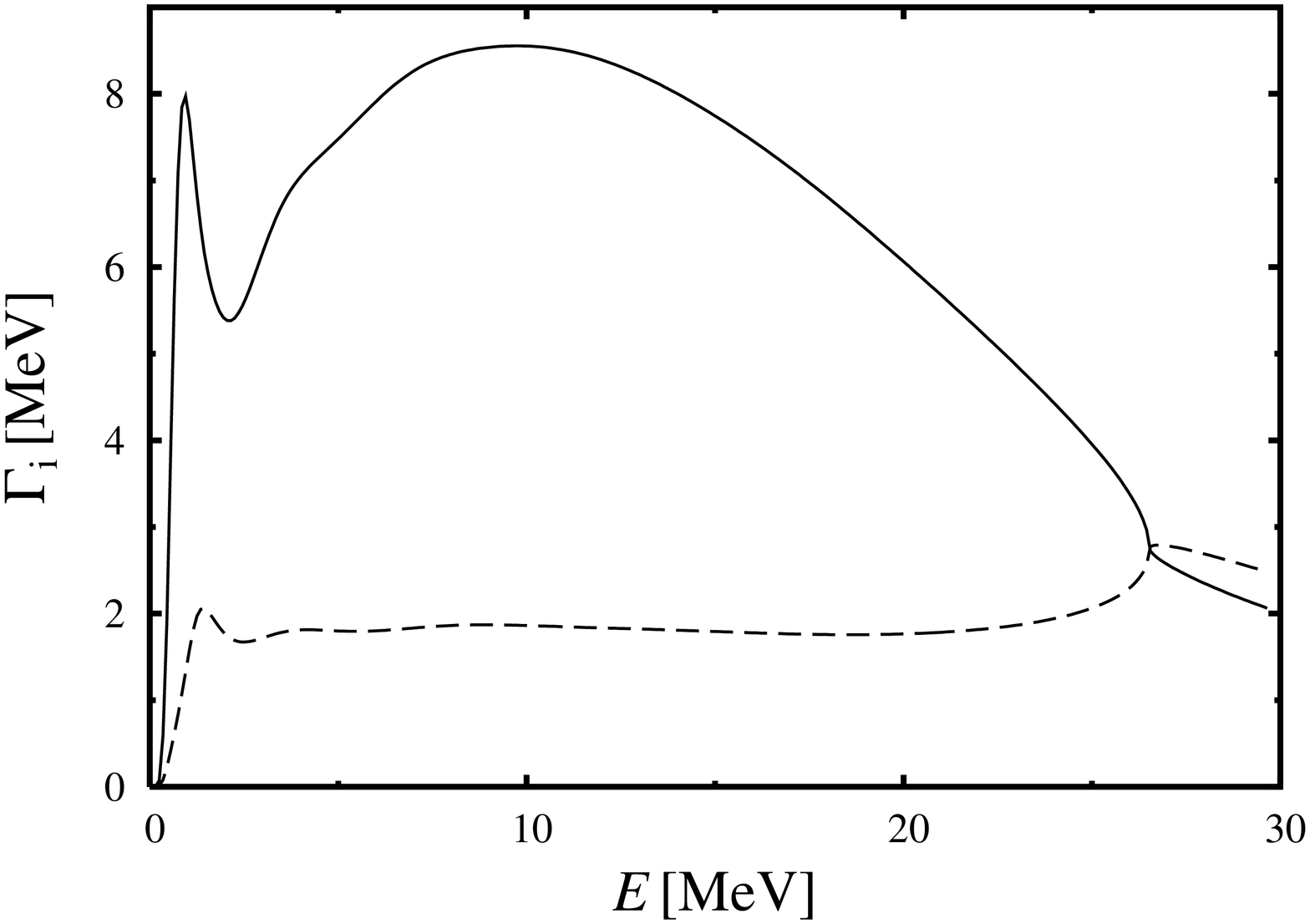}
\caption{Energy and width of lowest two eigenstates $0_1^+$, $0_2^+$ of an effective Hamiltonian 
(\ref{heff}) in $^{16}$Ne. These eigenstates form an exceptional point at $E=26.523\,\mbox{MeV}$. The continuum-coupling strength corresponding to this degeneracy is $V_0=-1182.32$ MeV$\cdot$fm$^3$.
Energies (in MeV) on the vertical ${\rm Re}E_i$-axis are plotted with respect to an arbitrarily chosen origin. Hence, only relative values of the real part of energies have a meaning in this representation. On the contrary, the width $\Gamma_i$ of each eigenstate depends only on a relative distance from the elastic threshold ($E=0$) and is not influenced by the convention chosen to plot ${\rm Re}E_i$.} 
\label{fig1}
\end{center}
\end{figure}
In this work, we study the one-proton continuum of $^{16}$Ne to look for a possible EP with 
$J^{\pi}=0^+$  in a physically relevant region of excitation energies and parameters of the SMEC. For the SM Hamiltonian ($H_{QQ}$ in (\ref{heff})) we take the ZBM Hamiltonian \cite{ZBM}. The residual coupling $H_{QP}$ between $Q$-subspace and the embedding continuum is given by: 
$V_{12}=V_0\delta(r_1-r_2)$. The true degeneracy of two $0^+$ resonances is searched for by varying the strength parameter $V_0$ in $H_{QP}$ and the total energy $E$ of $^{16}$Ne.  $E=0$ is chosen at the lowest one-proton emission threshold. In present studies, we consider the coupling of $0^+$ SM states to three lowest decay channels in $^{16}$Ne: 
$\Big{[}{^{15}{\rm F}}(1/2_1^+)\otimes{\rm p}_E(1s_{1/2})\Big{]}^{{0}^{+}}$ (the elastic channel),
$\Big{[}{^{15}{\rm F}}(5/2_1^+)\otimes{\rm p}_E(0d_{5/2})\Big{]}^{{0}^{+}}$, and
$\Big{[}{^{15}{\rm F}}(1/2_1^-)\otimes{\rm p}_E(0p_{1/2})\Big{]}^{{0}^{+}}$.
The energy dependence of $H_{QQ}^{\mbox{\small eff}}$ is due to the coupling of discrete states of $^{16}$Ne to the decay channels $\Big{[}{^{15}{\rm F}}(K^{\pi'})\otimes{\rm p}_E(l_{j})\Big{]}^{{J}^{\pi''}}$. This coupling depends explicitly on $E$. For $E>0$, $H_{QQ}^{\mbox{\small eff}}$  is complex-symmetric and its eigenvalues are complex. 
 
\begin{figure}[hbt]
\begin{center}
\includegraphics[width=6cm,angle=-90]{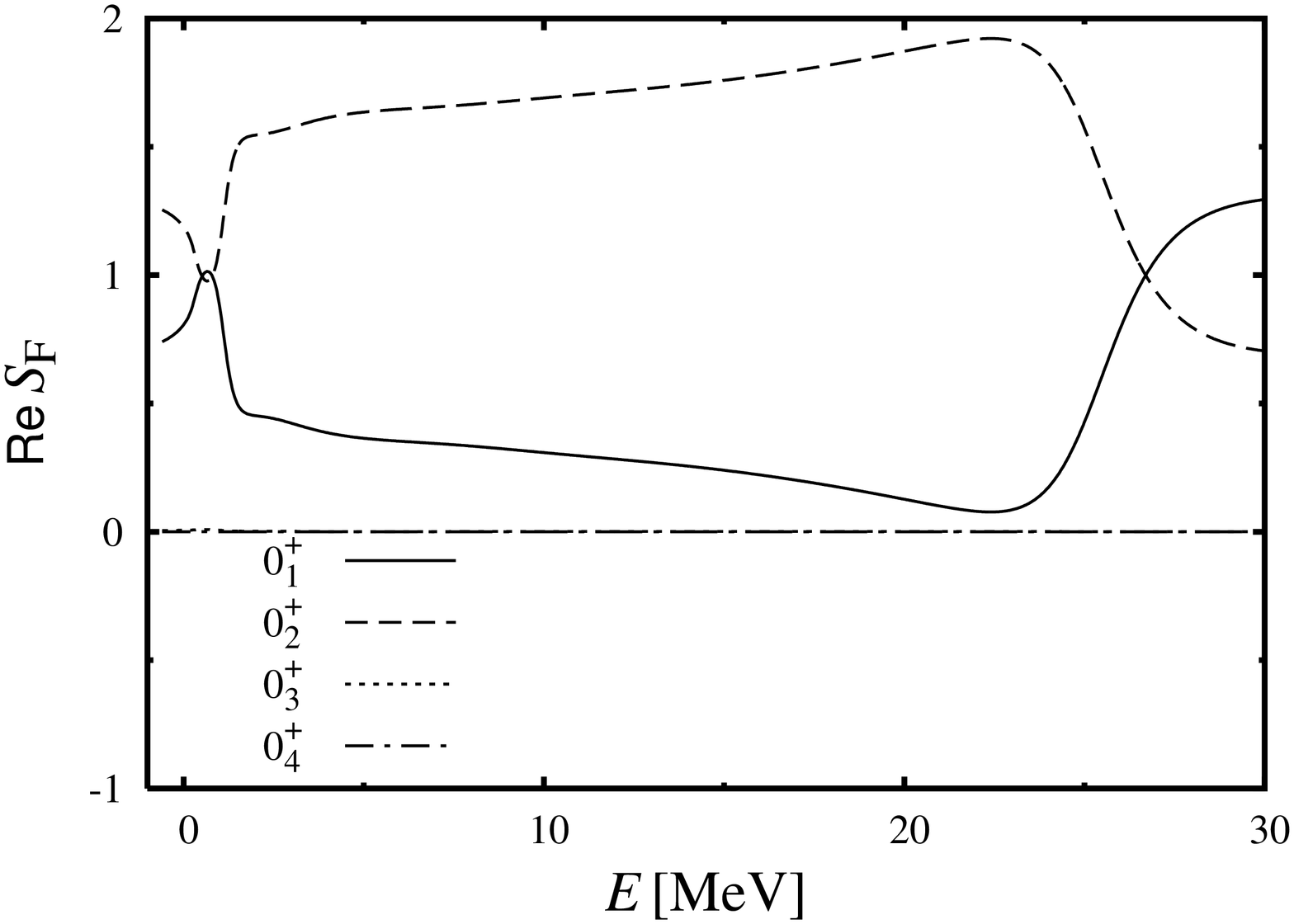}
\includegraphics[width=6cm,angle=-90]{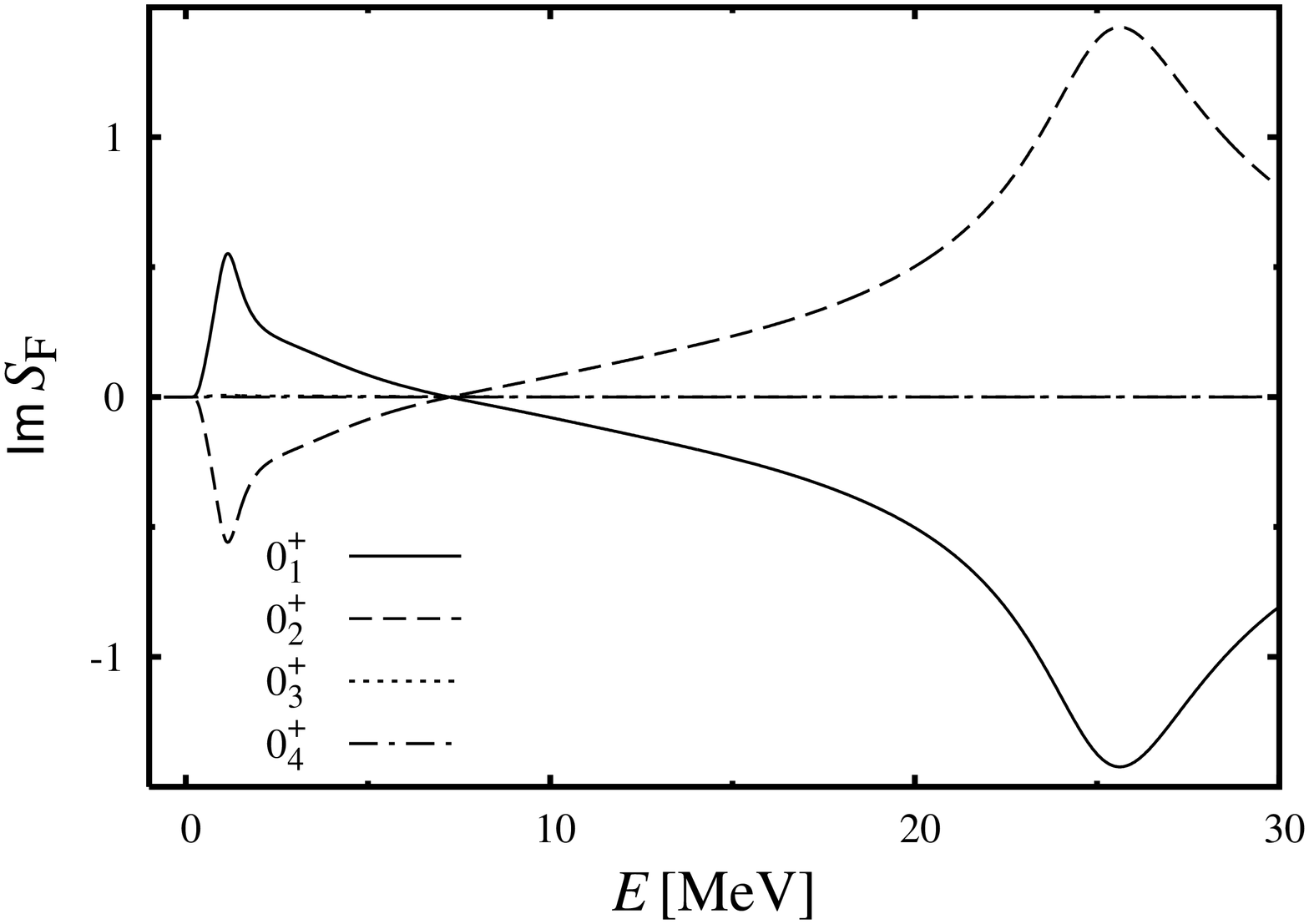}
\caption{Real and imaginary parts of the one-proton spectroscopic factor 
$\langle {^{16}{\rm Ne}(0_n^+)}|[{^{15}{\rm F}(1/2_1^+)\otimes{\rm p}(1s_{1/2})]^{^+}}\rangle$ 
to the ground state ($J^{\pi}=1/2_1^+$) of $^{15}$F in different $0_n^+$
states (${n=1,\cdots,4}$) of $^{16}$Ne ($V_0=-1100$ MeV$\cdot$fm$^3$).} 
\label{fig2}
\end{center}
\end{figure}
\begin{figure}[hbt]
\begin{center}
\includegraphics[width=6cm,angle=-90]{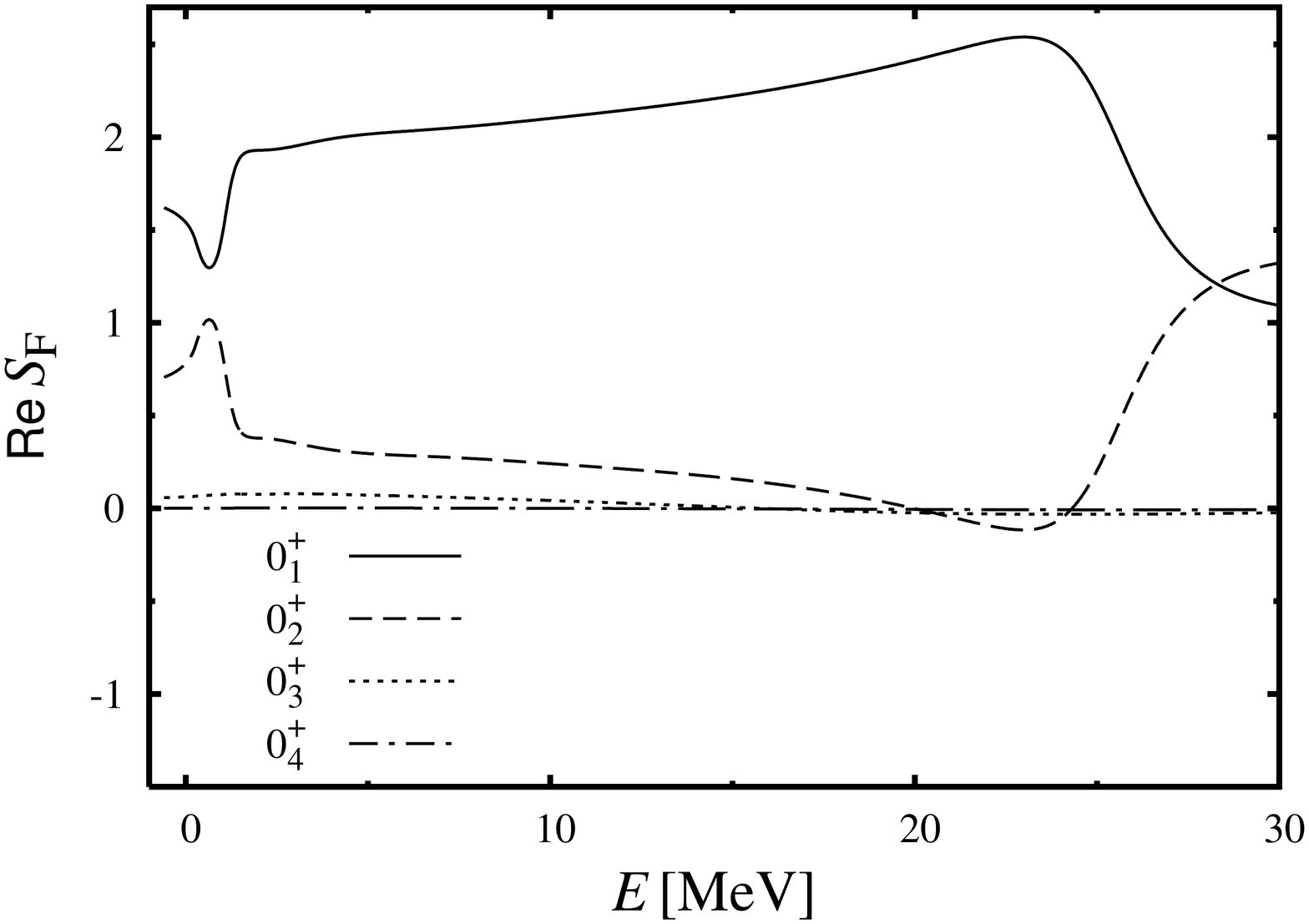}
\includegraphics[width=6cm,angle=-90]{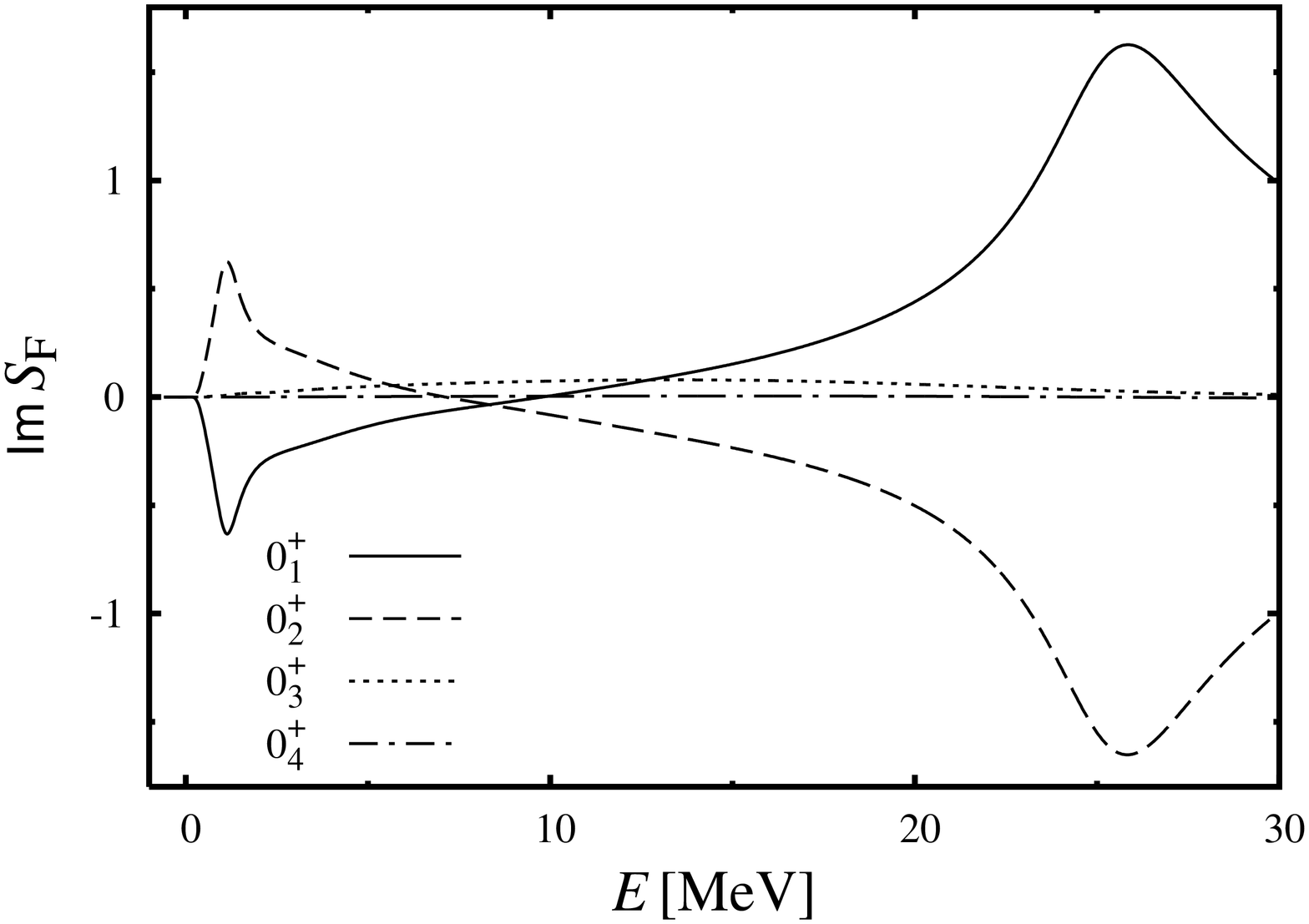}
\caption{Real and imaginary parts of the one-proton spectroscopic factor 
$\langle {^{16}{\rm Ne}(0_n^+)}|[{^{15}{\rm F}(5/2_1^+)\otimes{\rm p}(0d_{5/2})]^{^+}}\rangle$ 
to the first excited state ($J^{\pi}=5/2_1^+$)  of $^{15}$F
in different $0_n^+$ states (${n=1,\cdots,4}$) of $^{16}$Ne ($V_0=-1100$ MeV$\cdot$fm$^3$).} 
\label{fig3}
\end{center}
\end{figure}
\begin{figure}[hbt]
\begin{center}
\includegraphics[width=6cm,angle=-90]{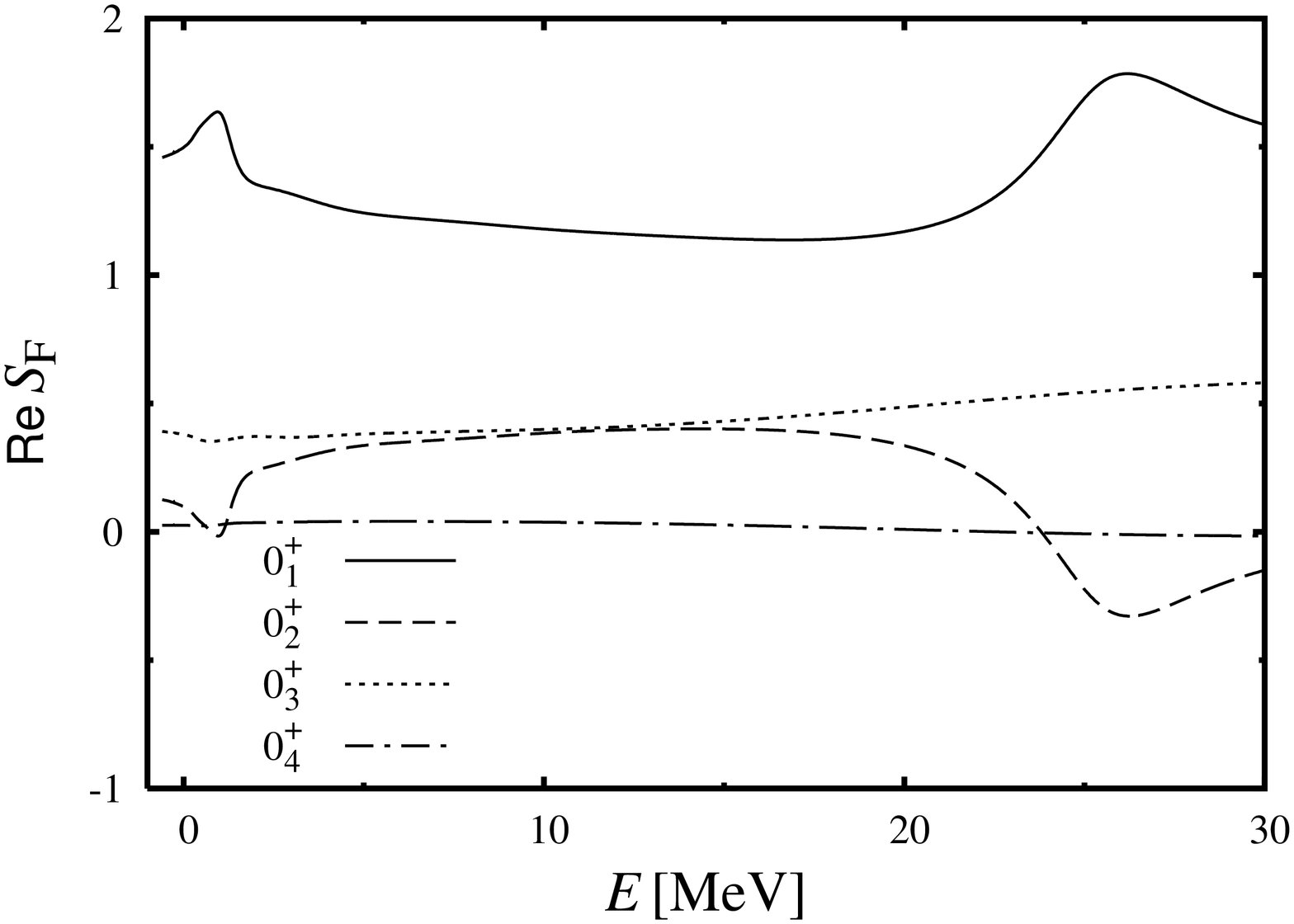}
\includegraphics[width=6cm,angle=-90]{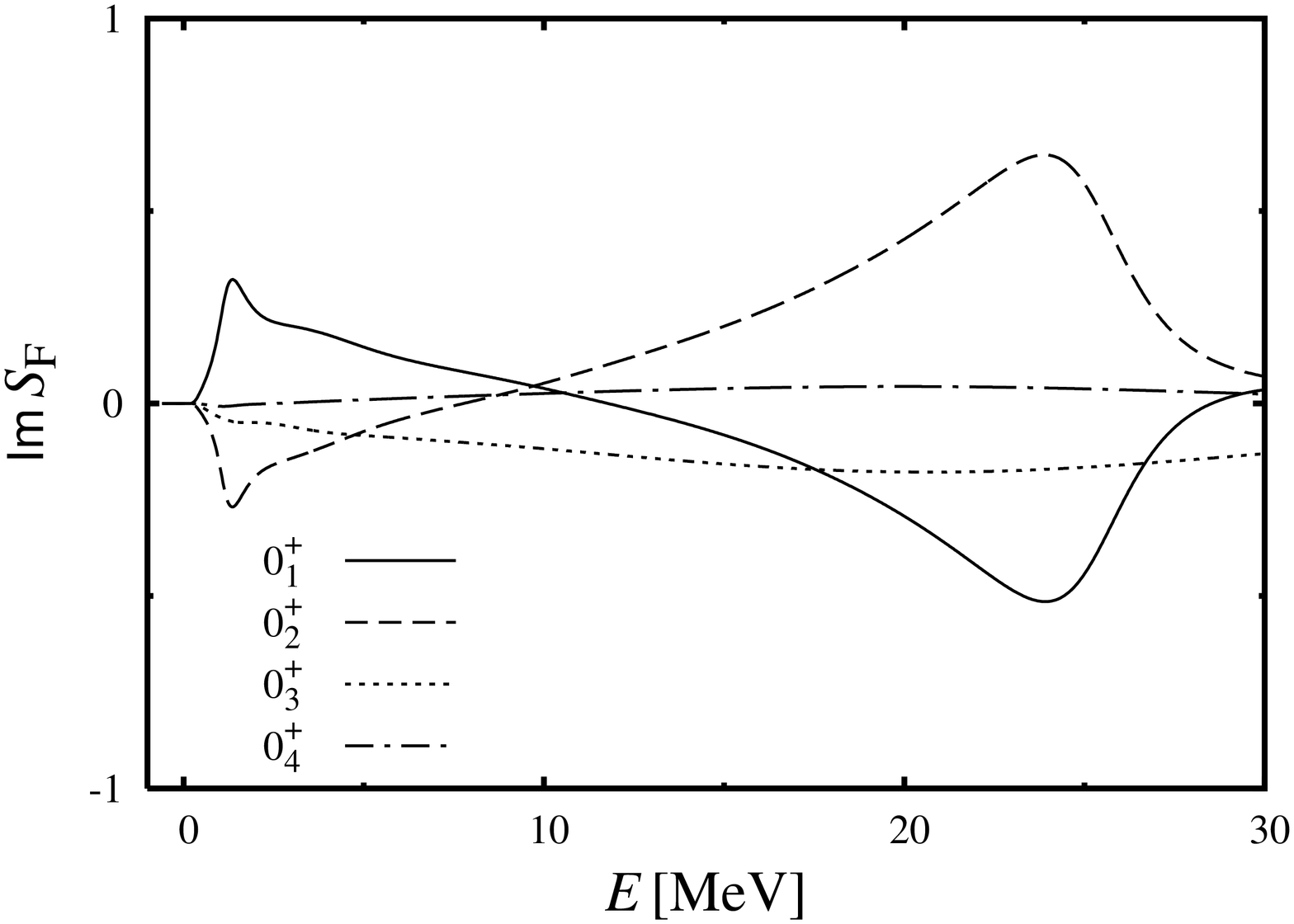}
\caption{Real and imaginary parts of the one-proton spectroscopic factor 
$\langle {^{16}{\rm Ne}(0_n^+)}|[{^{15}{\rm F}(1/2_1^-)\otimes{\rm p}(0p_{1/2})]^{^+}}\rangle$ 
to the second excited state ($J^{\pi}=1/2_1^-$)  of $^{15}$F
in different $0_n^+$ states (${n=1,\cdots,4}$) of $^{16}$Ne ($V_0=-1100$ MeV$\cdot$fm$^3$).} 
\label{fig4}
\end{center}
\end{figure}
Fig. \ref{fig1} shows energies and widths of $0_1^+$, $0_2^+$ eigenvalues as a function of the total energy of $^{16}$Ne for a fixed value of the  strength  $V_0=-1182.32$ MeV$\cdot$fm$^3$ of the residual coupling $H_{QP}$.  $0_1^+$, $0_2^+$ eigenstates and eigenvectors of $H_{QQ}^{\mbox{\small eff}}(E)$ coalesce at $E=26.523\,\mbox{MeV}$, forming an EP. Encircling this singularity in a parameter space of  $H_{QQ}^{\mbox{\small eff}}$, reveals a topological phase in $0_1^+$, $0_2^+$ eigenvectors. One can see a strong variation of $0_1^+$ and $0_2^+$ eigenvalues in the vicinity of thresholds for channels $\Big{[}{^{15}{\rm F}}(1/2^+)\otimes{\rm p}(1s_{1/2})\Big{]}^{0^+}$  
and  $\Big{[}{^{15}{\rm F}}(5/2^+)\otimes{\rm p}(0d_{5/2})\Big{]}^{0^+}$, respectively. The Coulomb interaction shifts the near-threshold minima of Re$E_i(E)$ slightly above the threshold in
channels $\Big{[}{^{15}{\rm F}}(1/2^+)\otimes{\rm p}(1s_{1/2})\Big{]}^{0^+}$  ($E=0$) for $0_1^+$
and  $\Big{[}{^{15}{\rm F}}(5/2^+)\otimes{\rm p}(0d_{5/2})\Big{]}^{0^+}$ ($E=0.671\,\mbox{MeV}$) for $0_2^+$. One should also notice that the energy $E$ of the lowest EP with $J^{\pi}=0^+$ does not necessarily correspond to the resonance energy which in SMEC is determined by solving a 
fixed-point equation for eigenvalues $E_i(E)$ of the effective Hamiltonian 
$H_{QQ}^{\mbox{\small eff}}(E)$ \cite{Oko03}. Real part of energies in Fig. \ref{fig1} are plotted with respect to an arbitrarily chosen zero on the axis ${\rm Re}E_i$ and only relative differences are meaningful. In the case considered in Fig. \ref{fig1}, if the ground state energy with respect to the elastic threshold ($E=0$) is fixed at $E_{0_1^+}=-0.5$ MeV or $+0.05$ MeV, then the energy of the $0_2^+$ state obtained from the fixed-point equation becomes $E_{0_2^+}=4.918$ MeV or 6.251 MeV, respectively.

One should stress an essential role of non-hermiticity of $H_{QQ}^{\mbox{\small eff}}$ in this problem. The hermitian Hamiltonian can be always diagonalized and eigenvalues, even if degenerate, always correspond to distinct eigenvectors. On the contrary, in the non-hermitian Hamiltonian one may find non-trivial Jordan blocks of eigenfunctions and singular points in a parameter space of the Hamiltonian where both eigenvalues and eigenvectors coalesce. These EPs have drastic effects on systems behavior, as we shall examine below on the example of spectroscopic factors.

Fig.~\ref{fig2} shows real and imaginary parts of the one-proton spectroscopic factor 
$\langle {^{16}{\rm Ne}(0_n^+)}|[{^{15}{\rm F}(1/2_1^+)\otimes{\rm p}(1s_{1/2})]^{^+}}\rangle$ for the four lowest $0^+$ states in $^{16}$Ne. This spectroscopic factor comes from a coupling of $0_n^+$ states 
(${n=1,\cdots,4}$) of $^{16}$Ne to the ground state $1/2_1^+$ of $^{15}$F.
 The strength of the residual $\delta$-force coupling between $Q$-subspace and the embedding continuum ($V_0=-1100$ MeV$\cdot$fm$^3$) is chosen to reproduce an experimental width 
 of $5/2_1^+$ state in $^{15}$F. 
Analogously, Fig.~\ref{fig3} and \ref{fig4} exhibit the 
$\langle {^{16}{\rm Ne}(0_n^+)}|[{^{15}{\rm F}(5/2_1^+)\otimes{\rm p}(0d_{5/2})]^{^+}}\rangle$ and
$\langle {^{16}{\rm Ne}(0_n^+)}|[{^{15}{\rm F}(1/2_1^-)\otimes{\rm p}(0p_{1/2})]^{^-}}\rangle$ spectroscopic factors, respectively. These spectroscopic factors result from a coupling to the first ($J^{\pi}=5/2_1^+$) and the second ($J^{\pi}=1/2_1^-$) excited states in the daughter nucleus $^{15}$F.
Imaginary part of the spectroscopic factor, which appears for energies above the first one-proton emission threshold ($E>0$), yields the uncertainty of a real part and is related to the decaying nature of considered $0^+$ states. For deeply bound states, an external mixing of SM states due to the coupling to common decay channels can be safely neglected and, hence, the spectroscopic factor calculated in SMEC becomes  close to the SM value. This tendency can be seen in Figs. \ref{fig2}-\ref{fig4} for $E<0$. 

Figs. \ref{fig2}-\ref{fig4} exhibit the external mixing of $J^{\pi}=0^+$ SM states  by showing the 
$E$-dependence of spectroscopic factors for corresponding $0^+$ SMEC eigenstates. This mixing is strongly dependent on the structure of individual $0^+$ SM states, the position of different one-proton emission thresholds, and the total energy $E$ of $^{16}$Ne. One can see in Figs. \ref{fig2}, \ref{fig3} that the one-proton spectroscopic factors are large for $0_1^+$ and $0_2^+$. The eigenstate $0_3^+$ has only a sizeable 
$\langle {^{16}{\rm Ne}(0_3^+)}|[{^{15}{\rm F}(1/2_1^-)\otimes{\rm p}(0p_{1/2})]^{^-}}\rangle$ spectroscopic factor.

One can notice two distinct phenomena in the $E$-dependence of spectroscopic factors: the threshold effect and the EP effect. The particle-emission threshold is a branch point in OQSs \cite{Baz69}. In Figs. \ref{fig2}-\ref{fig4}, one may notice a strong effect of the coupling to all three decay channels $\Big{[}{^{15}{\rm F}}(1/2_1^+)\otimes{\rm p}(1s_{1/2})\Big{]}^{0^+}$ (the elastic channel),
$\Big{[}{^{15}{\rm F}}(5/2_1^+)\otimes{\rm p}(0d_{5/2})\Big{]}^{0^+}$, and 
$\Big{[}{^{15}{\rm F}}(1/2_1^-)\otimes{\rm p}(0p_{1/2})\Big{]}^{0^-}$
for a pair of eigenstates $0_1^+$, $0_2^+$. Remaining $0^+$ states are effectively decoupled showing a weak $E$-dependence of spectroscopic factors in the vicinity of thresholds. The threshold effect is well localized in $E$ and leads to mutually compensating variations in spectroscopic factors for strongly coupled $0_1^+$ and $0_2^+$ eigenstates of  $H_{QQ}^{\mbox{\small eff}}$. These compensating variations of single-particle occupancies is a signature of an entanglement of the two externally coupled states in the scattering continuum. In other words, one cannot define any matrix element or expectation value of an operator which does not commute with the Hamiltonian on an individual state of the entangled couple of eigenstates. 

The effect of an EP at $E=26.523\,\mbox{MeV}$ leads to much stronger variations of all three spectroscpic factors: 
$\langle {^{16}{\rm Ne}(0_n^+)}|[{^{15}{\rm F}(1/2_1^+)\otimes{\rm p}(1s_{1/2})]^{^+}}\rangle$,
$\langle {^{16}{\rm Ne}(0_n^+)}|[{^{15}{\rm F}(5/2_1^+)\otimes{\rm p}(0d_{5/2})]^{^+}}\rangle$, and
$\langle {^{16}{\rm Ne}(0_n^+)}|[{^{15}{\rm F}(1/2_1^-)\otimes{\rm p}(0p_{1/2})]^{^-}}\rangle$ 
for $0_1^+$ and $0_2^+$ states. The same compensation mechanism as seen in the vicinity of the thresholds is found for spectroscopic factors in $0_1^+$ and $0_2^+$ states.  However, since the strength of the residual coupling ($V_0=-1100$ MeV$\cdot$fm$^3$) does not correspond  to an exact value ($V_0=-1182.32$ MeV$\cdot$fm$^3$) for which the true degeneracy of eigenvalues has been found, the variations seen are caused by the proximity of the EP and {\em not} by the EP itself. As a result, the EP effect is weaker in magnitude and less concentrated in energy $E$ than expected at an exact position of the EP. It is interesting to notice that the intricate entanglement of two continuum states caused by the EP is seen not only at the singularity point  in parameter space of the Hamiltonian
($E=26.523\,\mbox{MeV}, V_0=-1182.32$ MeV$\cdot$fm$^3$) but also relatively far away from it. In that sense, it is a robust phenomenon relatively weakly dependent on a fine-tuning of the Hamiltonian parameters and apt to be observed in nuclear physics experiments.

We acknowledge fruitful discussions with J. Dukelsky and W. Nazarewicz.

\end{document}